# Interfacial interaction in monolayer transition metal dichalcogenides ($MX_2$)/metal oxide heterostructures and its effects on electronic and optical properties: The case of $MX_2/CeO_2$

Ke Yang[1], Wei-Qing Huang[1*], Wangyu Hu[2], Gui-Fang Huang[1#], Shuangchun Wen[1]

[1] Department of Applied Physics, School of Physics and Electronics, Hunan University, Changsha 410082, China

[2] School of Materials Science and Engineering, Hunan University, Changsha 410082, China

Two-dimensional transition metal dichalcogenides ($MX_2$)/metal oxide heterostructures have shown unique physical properties, making them promising materials for various applications ranging from photocatalysis to solar energy conversion. Understanding the interfacial interactions is highly desirable for designing these heterostructures having excellent performance. Here we systematically study the interfacial interaction in monolayer $MX_2$ (M=Mo, W; X=S, Se)/$CeO_2$ heterostructures and its effects on electronic and optical properties by density functional theory. It is found that the interfacial interaction in the $MX_2/CeO_2$ depends predominantly on the chalcogen (X) element. Particularly, the band gap variation and important electronic states at conduction band minimum or valence band maximum of the heterostructures are determined by the strength of interfacial interaction. The $MX_2/CeO_2$ heterostructures with the same chalcogen (X) element have similar absorption spectra from ultraviolet to near-infrared regions. These results suggest that the chalcogen (X) element is a key factor in tuning the properties of $MX_2$/metal oxide heterostructures.

[*]. Corresponding author. *E-mail address:* wqhuang@hnu.edu.cn

[#]. Corresponding author. *E-mail address:* gfhuang@hnu.edu.cn

Two-dimensional transition metal dichalcogenides (2D TMDs), with the formula $MX_2$ (where M is a transition metal element and X is a chalcogen element), have attracted increasing attention in recent years because of their unique optoelectronic and catalytic properties caused by their tunable electronic structures.[1-3] The intra-layer M–X bonds in 2D TMDs are predominantly covalent in nature, whereas neighboring layers are interacted by weak van der Waals (vdWs) forces thus allowing them to readily cleave along the layer plane.[4,5] This bonding characteristic makes monolayer or few layers TMDs particularly attractive for integration of highly disparate materials to form vdWs heterojunction without the constraints of crystal lattice matching,[2] thus maintaining their excellent properties.[5]

Substantial efforts have recently been concentrated on building 2D TMDs-based heterostructures with optimized performance to meet the need for a variety of applications.[6-8] Among them, 2D TMDs-metal oxides (MOs) heterostructures attract particular interest and exhibit many excellent properties compared with their counterparts.[9-13] For examples, $MoS_2/TiO_2$ nanostructures not only possess enhanced photocatalytic activities due to band matching between two components,[9-13] but also have a good recyclability compared with pure $TiO_2$.[12] $MoS_2/ZnO$ nanocomposites exhibit superior optical properties,[14,15] such as enhanced nonlinear absorption and scattering.[16] $MoS_2/CuO$ heterojunctions with staggered type-II band alignment have improved photocatalytic performance and $H_2O$ adsorption.[17] With a gap of about 1.7 eV, monolayer $WS_2$ can sensitize $TiO_2$ toward longer wavelengths (~700 nm), thus showing high photocatalytic performance.[18-20] $WS_2/WO_3$ heterostructures has also demonstrated excellent synergistic effects which facilitate the kinetics of the hydrogen evolution reaction.[21] It is generally assumed that the enhanced properties of TMDs-MOs heterostructures is mainly due to the robust separation of photoexcited charge carriers at interface.[22-24] However, these experimental and theoretical investigations on the 2D TMDs-MOs heterostructures are in a scattered and isolated way, and lack of systematic research so far. The primary factors on the interfacial interactions and superior properties of 2D TMDs-MOs heterostructures are not clear. For instance, why do the $WS_2/TiO_2$ heterostructures exhibit much higher photoactivity compared to $MoS_2/TiO_2$ ones?[25] Unveiling the interfacial interactions and the underlying mechanism for enhanced properties of 2D TMDs-MOs heterostructures are desired, yet challenging.

Herein we systematically explore the interfacial interaction, and its influence on the electronic and optical properties of 2D $MX_2$ (M=Mo, W, X=S, Se) and $CeO_2$, as a representative metal oxides, using *ab initio* calculations. The choice of $CeO_2$ is owing to not only its important applications,[26] but also as one of the most appropriate substrates for 2D materials, such as graphene,[27] hexagonal BN,[28] and g-$C_3N_4$,[29] even more for precious metal single-atom Pt.[30] Most importantly, many experimental investigations on $MX_2/CeO_2$ heterostructures[31-33] provide reliable supports for theoretical research. The results demonstrate that the interfacial interaction in the $MX_2/CeO_2$ heterostructures depends predominantly on the chalcogen (X) element. The near-gap electronic structure of these heterostructures can be readily tuned by simply choosing the chalcogen (X) element. Furthermore, the $MX_2/CeO_2$ heterostructures with the same chalcogen (X) element have similar absorption spectra from ultraviolet to near-infrared regions. The chalcogen-dependence of the interfacial interaction, near-gap electronic structure, and optical properties is expected to be general in other 2D $MX_2$/MOs heterostructures.

The calculations are performed by using the Vienna ab initio simulation package (VASP)[34,35] based on density-functional theory (DFT) with the projector augmented wave (PAW) method.[36] Local density approximation (LDA) is adopted due to long-range vdWs interactions are expected to be significant in these systems. All of the calculations are performed by the DFT/LDA+U method (Ce 4*f* and O 2*p* are 9.0 and 4.5 eV) to get the correct band gap. The kinetic energy cutoff is 500 eV for the plane wave basis. Brillouin zone integrations are used on grids of $7 \times 7 \times 1$ Monkhorst−Pack k-points. In the geometrical optimization, total energy and all forces on atoms are converged to less than $10^{-6}$ eV and 0.03 eV/Å. To construct the periodic interface, we choose a ($\sqrt{3} \times \sqrt{3}$) stoichiometric cubic $CeO_2(111)$ surface slab (nine layers) containing 18 O atoms and 9 Ce atoms, among which the three bottom layers are fixed at the bulk position, matching a (2×2) monolayer $MX_2$ (4 M atoms and 8 X atoms). This gives rise to minor compressed deformation of $CeO_2(111)$, resulting into about 3.0 % lattice mismatch. The vacuum space is 15 Å, which is enough to separate the interaction between periodic images.

Interfacial interaction in the vdWs heterostructures can be properly assessed by the separation between constituents: the smaller the distance, the stronger the interaction.[27] For the fully relaxed geometries, the equilibrium distances, $d$, between monolayer $MX_2$ and $CeO_2$ (111) surface are calculated and listed in Table I. For the four vdWs heterostructures, the equilibrium distance, $d$, ranges from 2.76 to 2.91 Å, indicating that the interactions between monolayer $MX_2$ and $CeO_2$ (111) surface are indeed vdWs rather than covalent, in accordance with previous reports (2.96, 2.79, 3.32 and 3.01 Å for graphene/$CeO_2$,[27] $MoS_2$/$MoSe_2$,[5] and $MoS_2$/ZnO(0001),[23] respectively).

The key feature of the results in Table I is that the equilibrium distance, $d$, is predominately determined by the chalcogen (X) element in the $MX_2$/$CeO_2$ heterostructures. The $d$ in $MSe_2$/$CeO_2$(111) heterostructures are about 2.90 Å, larger than those in corresponding $MS_2$/$CeO_2$(111) ones, meaning that the interfacial interaction in $MX_2$/$CeO_2$ is weaken as the atomic number of chalcogen (X) elements increases from the top down. In contrast, the effect of group 6 transition metal (M) element on the $d$ is opposite, although it is much weaker than the chalcogen (X) element. Therefore, the interfacial interaction in $MX_2$/$CeO_2$ is mainly dependent on the chalcogen (X) element. This dependence can be elucidated by considering that the monolayer $MX_2$ is a structure of tri-layer X-M-X sheet. In 2D TMDs-MOs heterostructures, the distances between the chalcogen (X) elements facing to the MOs and the surface of MOs are smaller than 3.0 Å, whereas those between the transition metal (M) elements and the surface of MOs are generally bigger than 4.3 Å. Such distance difference directly leads to the interfacial interaction mainly determined by the chalcogen (X) elements rather than the transition metal (M) elements in 2D TMDs-MOs heterostructures.

The stability of $MX_2$/$CeO_2$(111) heterostructures can be evaluated by the interfacial adhesion energy, according to the following equation:

$$E_{ad} = E_{Comb} - E_{CeO_2(111)} - E_{MX_2} \qquad (1)$$

where $E_{Comb}$, $E_{CeO_2(111)}$, and $E_{MX_2}$ represent the total energy of the relaxed $MX_2$/$CeO_2$(111) heterostructures, pure $CeO_2$(111) surface and monolayer $MX_2$, respectively. With this definition, more negative $E_{ad}$ predicts that the adsorption is more stable. The calculated $E_{ad}$ of the four vdWs heterostructures are negative (Table I), revealing a rather strong interaction between monolayer $MX_2$ and $CeO_2$(111) surface, and the high thermodynamically stability of these systems. Compared with MSe$_2$/CeO$_2$(111), MS$_2$/CeO$_2$(111) heterostructures have more negative $E_{ad}$, meaning having more stronger interfacial interaction. This result again demonstrates that the interfacial interaction in MX$_2$/CeO$_2$ heterostructures relies mostly on the chalcogen (X) element.

Before exploring the effect of interfacial interaction on the electronic properties of MX$_2$/CeO$_2$(111), let us here briefly review the band structures of monolayer MX$_2$ and CeO$_2$. The calculated band gaps of monolayer MS$_2$ (1.86, 1.99 eV for MoS$_2$ and WS$_2$) are respectively bigger than those of MSe$_2$ (1.64, 1.68 eV for MoSe$_2$ and WSe$_2$), which agree well with previous theoretical studies.[4,38] For pure MX$_2$, its conduction band (CB) bottom is mainly constituted of M $d$ and X $p$ states, while the valence band (VB) top is composed of M $d$ states. This maybe one of the most important factors for the lower photocatalytic properties of pure MX$_2$. Bulk CeO$_2$ is a indirect band gap semiconductor with a large band gap of 3.2 eV,[38] while the CeO$_2$(111) surface is a direct band gap one and its band gap decreasing to 3.1 eV.[27] The discrepancy of vdW interfacial interactions in MX$_2$/CeO$_2$ heterostructures can also be illuminated by comparing their band structures, as shown in Fig. 1. The electronic structures of both the monolayer MX$_2$ and the CeO$_2$(111) surface are well-preserved (The band structures of pure monolayer MX$_2$ and CeO$_2$(111) are not given here). Interestingly, the CB bottom and VB top of the MX$_2$/CeO$_2$ heterostructures seem to be composed of those of MX$_2$, except the CB bottom of WSe$_2$/CeO$_2$(111) which are constituted by those of CeO$_2$(111) surface (Fig. 1(d)). However, the vdW interaction will lead to the electron-hole wave function overlap between MX$_2$ and CeO$_2$, which can be directly visualized by the electron and hole density distributions of the conduction band minimum (CBM) and valence band maximum (VBM) states, as displayed in Fig. 2. For the MS$_2$/CeO$_2$ heterostructures, the VBM states are formed from the O 2$p$, S 3$p$ and M $d_z^2$ (Mo 4$d_z^2$, W 5$d_z^2$) states due to interfacial interaction, whereas the CBM states are only composed of M $d$ ($d_{xy}$, $d_z^2$, and $d_{x^2-y^2}$) states. In contrast, the VBM states arise only from the M $d$ ($d_{xy}$, and $d_{x^2-y^2}$) orbitals, whereas the CBM states are the hybridization of M $d$ ($d_{xy}$, $d_z^2$, and $d_{x^2-y^2}$) and Ce 4$f$ states in the MSe$_2$/CeO$_2$ heterostructures. This indicates that the band edges of 2D MX$_2$/CeO$_2$ heterostructures can be tuned by choosing the chalcogen (X) element. This dependency is very important for different applications of this kind of heterostructures. Under illumination, for example, the electrons (at VBM) in the CeO$_2$ (2D MSe$_2$) will be directly excited to the MS$_2$ (CeO$_2$), thus not only facilitating the spatial separation of electron–hole pairs, but also enhancing the reducing (oxidizing) capability of the $MS_2$ ($MSe_2$) in the 2D $MX_2/CeO_2$ heterostructures.

Interfacial interaction leads to the reduction of band gap of the 2D $MX_2/CeO_2$ heterostructures compared with their corresponding constituents. Unexpectedly, the amount of decrease in their band gaps also relys on the chalcogen (X) element: Compared pure monolayer $MX_2$, the band gaps of $MS_2/CeO_2(111)$ are dropped by about 0.3 eV, whereas those of $MSe_2/CeO_2(111)$ are reduced by only 0.04 eV approximately, as listed in Table I. This is because that the interfacial interaction in the $MS_2/CeO_2(111)$ is stronger than that in $MSe_2/CeO_2(111)$, in accordance with the smaller interfacial separation and adhesion energy of the former. The reduced band gap makes the 2D $MX_2/CeO_2$ heterostructures absorb more sunlight and improve the utilization of solar energy. Another outstanding characteristic in Fig. 1 is the influence of chalcogen (X) element on the positions of VBM and CBM, although the four heterostructures are all indirect band gap semiconductors. When the chalcogen (X) is S element, the stronger interaction in the $MS_2/CeO_2(111)$ makes a large shift of the energy levels at G-point, resulting into the G-point states are higher than the K-point states in energy (Figs. 1(a) and (c)). The upward shift of G-point state in the $MS_2/CeO_2(111)$ is due to the hybridization of the O 2$p$ and S 3$p$, M $d_z^2$ (Mo 4$d_z^2$, W 5$d_z^2$) states, among them the O 2$p$ orbital has higher energy and elevates the VBM. The weaker interaction in the $MSe_2/CeO_2(111)$ has not obviously impact on the band edge: the CBM and VBM of $MSe_2/CeO_2(111)$ remain at near K-point, and are closer together along K-G. Similar results has also been reported in $WS_2/MoS_2$ heterostructure.[37] Based on the results given above, the interfacial interaction and near-gap electronic structure of the 2D $MX_2/CeO_2$ heterostructures are demonstrated to be relying overwhelmingly on the chalcogen (X) element.

The variation of near-gap electronic structure of the 2D $MX_2/CeO_2$ heterostructures implies a substantial charge transfer between the involved constituents. In order to visualize the charge transfer at the interface, three-dimensional charge density difference have been calculated according to the following equation:

$$\Delta \rho = \rho_{MX_2/CeO_2(111)} - \rho_{MX_2} - \rho_{CeO_2(111)} \qquad (2)$$

where $\rho_{MX_2/CeO_2(111)}$, $\rho_{MX_2}$ and $\rho_{CeO_2(111)}$ are the charge densities of the heterostructures, monolayer $MX_2$ and $CeO_2(111)$ surface in the same configuration, respectively. Fig. 4 shows the charge density difference in the four heterostructures, where the yellow and cyan regions represent charge accumulation and depletion in the space, respectively. Clearly, the charge redistribution mainly occurs at the interface of $MX_2/CeO_2(111)$ heterostructures, while there is almost no charge transfer to the $CeO_2$ farther from the interface. The charge accumulation and depletion regions appear alternately: the former appears at the region where the X atom over the O atom (marked by blue half-ellipse), whereas the latter emerges next (marked by red ellipse) in the $WX_2/CeO_2(111)$ and $MoSe_2/CeO_2(111)$ heterostructures (see Figs. 3(b-d)). On the contrary, the alternant distribution of charge accumulation and depletion regions in the $MoS_2/CeO_2(111)$ heterostructure is opposite (Fig. 3(a)). This is similar to the case of g-$C_3N_4$/Graphene.[39] To quantitatively analyze the charge transfer at the interface, the effective net charge from one constituent to another in these heterostructures can be analyzed on the basis of the Bader method,[40] and the results are listed in Table I. Apparently, the chalcogen (X) element determines the amount of the transferred charge in the $MX_2/CeO_2(111)$ heterostructures. The amount of electrons transferred from $CeO_2$ to monolayer $MS_2$ is larger than that from $CeO_2$ to monolayer $MSe_2$: 0.029 (0.028) electron transfers from $MoS_2$ ($WS_2$) to $CeO_2$, whereas only 0.007 (0.002) electron from $MoSe_2$ ($WSe_2$) to $CeO_2$, which further demonstrates the interfacial interaction are stronger in $MS_2/CeO_2$ than $MSe_2/CeO_2$ heterostructures. To expound the origin of such an interface electron transfer in these heterostructures, we have calculated their work functions by aligning the Fermi level relative to the vacuum energy level. They are calculated to be 5.36, 4.89, 5.21, 4.51 and 6.03 eV for monolayer $MoS_2$, $MoSe_2$, $WS_2$, $WSe_2$ and the $CeO_2(111)$ surface, respectively. The spontaneous interfacial charge transfer in these heterostructures can be simply rationalized in terms of the difference on their work functions.

Experiments have demonstrated that coupling $MX_2$ nanosheets can enhance the visible adsorption and photocatalytic activity of metal oxides photocatalysts.[31,33] To explore the effects of the chalcogen (X) element on the optical properties of the 2D TMDs-MOs heterostructures, their frequency-dependent dielectric matrixes are calculated by the Fermi golden rule within the dipole approximation. The imaginary part $\varepsilon_2$ of the dielectric function $\varepsilon$ is calculated from the momentum matrix elements between the occupied and unoccupied wave functions, given by:

$$\varepsilon_2 = \frac{ve^2}{2\pi\hbar m^2\omega^2} \int d^3k \sum_{n,n'} |\langle kn|p|kn'\rangle|^2 f(kn)(1-f(kn'))\delta(E_{kn} - E_{kn'} - \hbar\omega) \quad (3)$$

where $\hbar\omega$ is the energy of the incident photon, p is the momentum operator $(\hbar/i)(\partial/\partial x)$, $(|kn\rangle)$ is a crystal wave function and $f(kn)$ is Fermi function. The real part $\varepsilon_1$ of the dielectric function $\varepsilon$ is evaluated from the imaginary part $\varepsilon_2$ by Kramer–Kronig transformation. The absorption coefficient $I(\omega)$ can be derived from $\varepsilon_1$ and $\varepsilon_2$, as following:

$$I(\omega) = \sqrt{2}\omega \left[\sqrt{\varepsilon_1^2(\omega) + \varepsilon_2^2(\omega)} - \varepsilon_1(\omega)\right]^{1/2} \quad (4)$$

which depends on $\varepsilon_1$ and $\varepsilon_2$ and thus on the energy. Taking into account the tensor nature of the dielectric function, $\varepsilon_1(\omega)$ and $\varepsilon_2(\omega)$ are averaged over three polarization vectors (along x, y, and z directions). All other optical constants can also be obtained. Generally, the optical absorption property of a semiconductor is closely related to its electronic band structure, which is a very important factor to determinate the photocatalytic activity.

The calculated absorption spectra of $MX_2/CeO_2(111)$ heterostructures are illustrated in Fig. 4. For the $CeO_2(111)$ surface, the optical absorption occurs at about 400 nm, which is attributed to the intrinsic transition from the O 2*p* to Ce 4*f* orbitals (~3.1 eV). Significantly, the $MX_2/CeO_2(111)$ heterostructures have a large redshift of the absorption edge and strong absorption in both ultraviolet and entire visible spectral regions which make them more suitable for use as optical detector or photocatalyst than do pure $CeO_2$ or monolayer $MX_2$. The curves of the four heterostructures in Fig. 4 exhibit different behaviors, although they all reflect the feature of the indirect bandgap semiconductors. Interestingly, the absorption curves of $MS_2/CeO_2(111)$ have similar trend, and so is the $MSe_2/CeO_2(111)$, indicating that the shape of absorption spectrum of $MX_2/CeO_2(111)$ is also dependent on the chalcogen (X) element. Compared to others, the absorption of $MoS_2/CeO_2(111)$ and $WS_2/CeO_2(111)$ heterostructures are much stronger from 370 to 420 nm, whereas those of $MoSe_2/CeO_2(111)$ and $WSe_2/CeO_2(111)$ ones are much higher from 490 to 580 nm. The strong resonant-like absorption peaks (about 430, 500, and 590 nm for $MoS_2/CeO_2(111)$, $MoSe_2/CeO_2(111)$, and $WS_2/CeO_2(111)$ heterostructures, respectively) in the absorption spectra are closely related to their respective unique near-gap electronic structures.

We now discuss the mechanism of enhanced optical absorption and improved photocatalytic activity of the $MoS_2/CeO_2$ heterostructure.[31-33] For the $MoS_2/CeO_2$ heterostructures, firstly, the levels at CB bottom and VB top of $MoS_2$ are embedded into the band gap of $CeO_2$, and furthermore, these levels hybridize with the orbitals from $CeO_2$, resulting into a smaller band gap compared to their individuals. The small band gap (~1.5 eV) enables the $MoS_2/CeO_2$ heterostructures to absorb more light in not only ultraviolet but also entire visible spectral regions (Fig. 4), which is in agreement with experimental results.[31] Secondly, the electrons in $CeO_2$ (i.e., at VBM) can be directly excited to $MoS_2$, resulting into the robust separation of photoexcited charge carriers between them in photocatalysis. Finally, some charged W (Mo) atoms at basal planes, initially catalytically inert, will turn out to be active sites due to charge transfer, making the monolayer $MoS_2$ to be a highly active co-catalyst in the heterostructures. Monolayer $MX_2$ (such as $MoS_2$) acting as co-catalyst in their heterostructures has recently been demonstrated.[24]

In summary, the electronic structure, interfacial charge transfer and optical properties of 2D $MX_2/CeO_2$ heterostructures have been investigated using DFT calculations to explore the interfacial interaction and its effects. It is revealed that the interfacial interaction between monolayer $MS_2$ and $CeO_2(111)$ surface is much stronger than that between $MSe_2$ and $CeO_2(111)$ surface. The strength of interfacial interaction largely determines the variation of band gap, near-gap electronic structure, and optical properties of 2D $MX_2/CeO_2$ heterostructures. The chalcogen-dependence would be general in other 2D $MX_2/MOs$ heterostructures. This work suggests that choosing appropriate chalcogen (X) element is an effective strategy to control the electronic structure and other properties of the 2D $MX_2/MOs$ heterostructures.

## Acknowledgements

This work is supported by the National Natural Science Foundation of China (Grant Nos. 11574079).

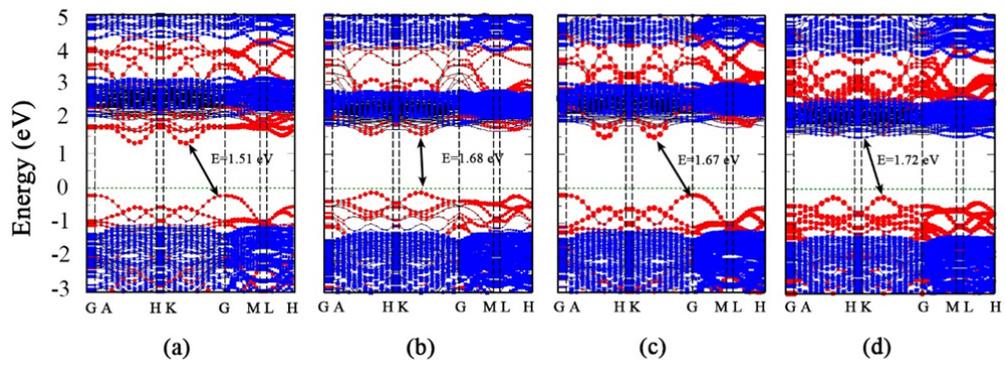

Figure 1. Band structure: (a) $MoS_2/CeO_2(111)$, (b) $MoSe_2/CeO_2(111)$, (c) $WS_2/CeO_2(111)$ and (d) $WSe_2/CeO_2(111)$. Red and blue represent $MX_2$ and $CeO_2(111)$, respectively. The Fermi level (dashed lines) is set to zero energy.

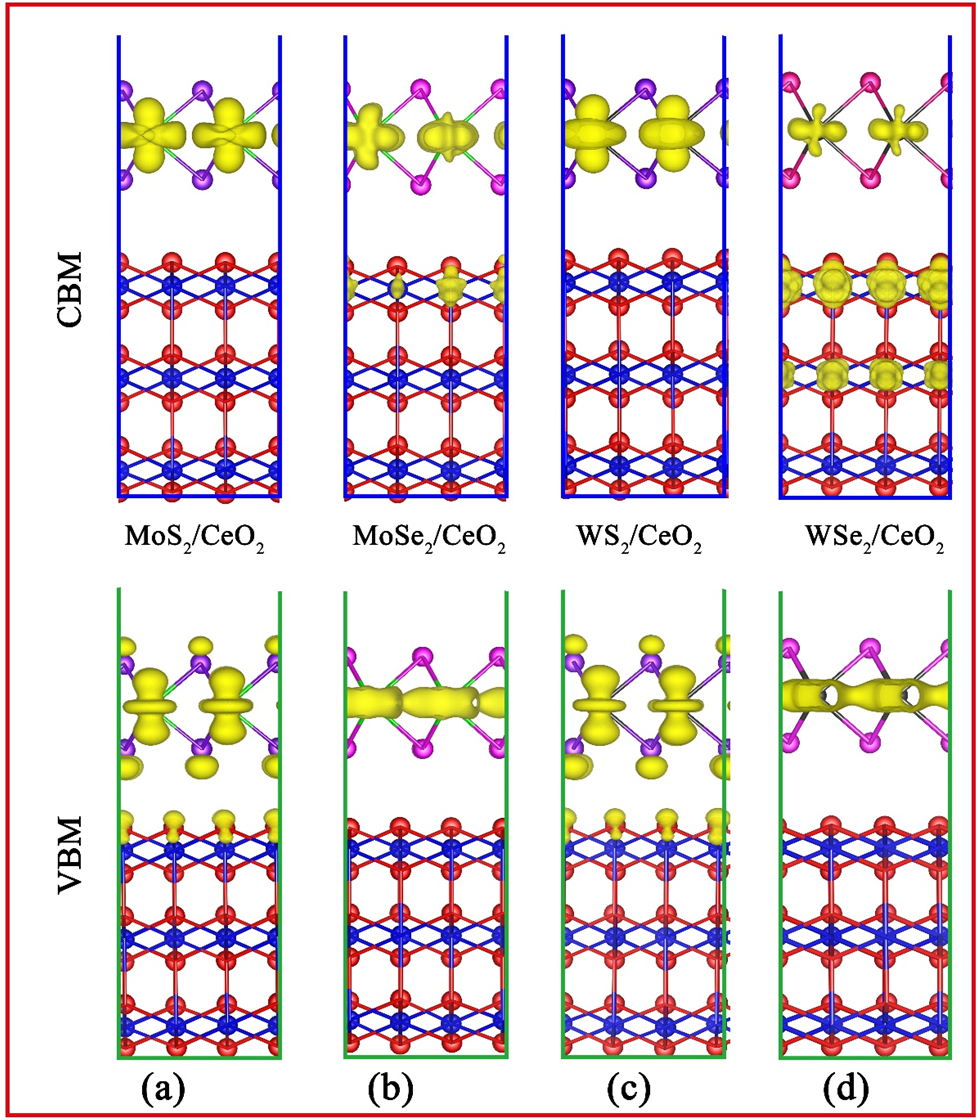

Figure 2. Maps of the electron and hole density distributions with an isovalue 0.004 e/Å$^3$ for CBM and VBM: (a) MoS$_2$/CeO$_2$(111), (b) MoSe$_2$/CeO$_2$(111), (c) WS$_2$/CeO$_2$(111) and (d) WSe$_2$/CeO$_2$(111). Blue, red, purple, pink, green and black spheres denote Ce, O, S, Se, Mo and W atoms, respectively.

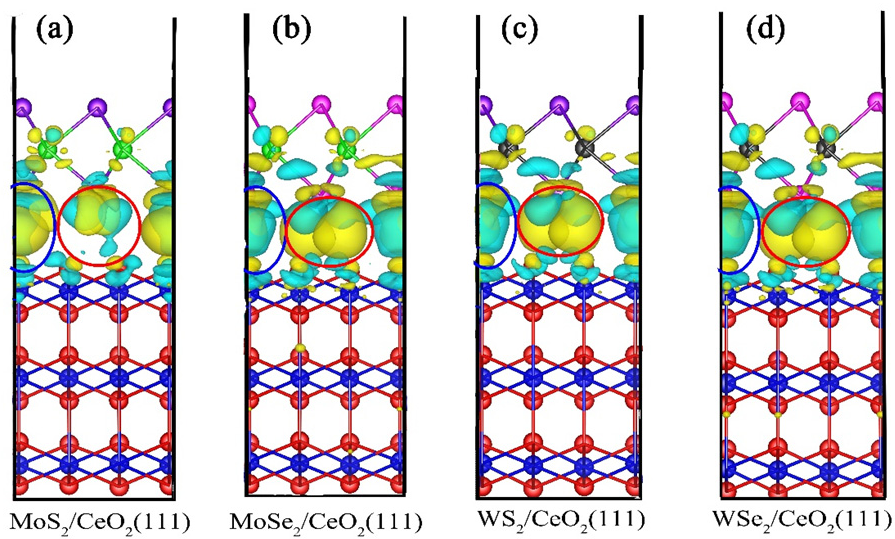

Figure 3. Three-dimensional charge density differences for (a) $MoS_2/CeO_2(111)$, (b) $MoSe_2/CeO_2(111)$, (c) $WS_2/CeO_2(111)$ and (d) $WSe_2/CeO_2(111)$. The yellow and cyan regions denote charge accumulation and charge depletion, respectively; the isosurface value is 0.0004 e/Å$^3$. Blue, red, purple, pink, green and black spheres denote Ce, O, S, Se, Mo and W atoms, respectively.

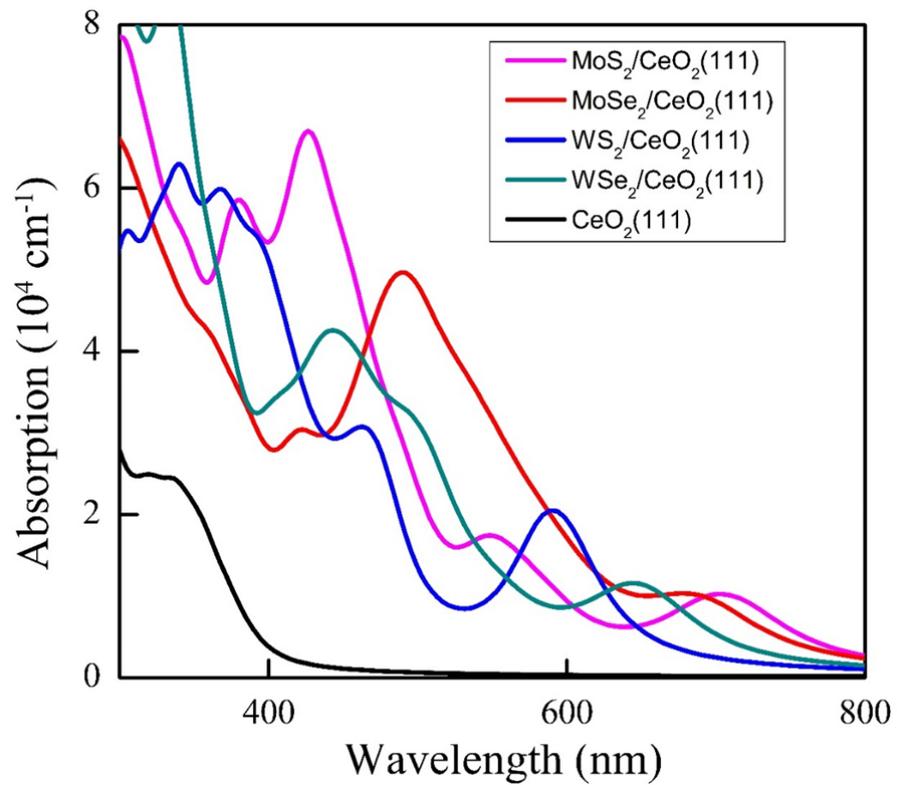

Figure 4. Calculated absorption spectra of the pure CeO$_2$(111) surface and MX$_2$/CeO$_2$(111) heterostructures for the polarization vector perpendicular to the surface.

TABLE I. Interfacial distance (d), adhesion energy ($E_{ad}$), bandgap ($E_{gap}$), and Bader charge analysis of optimized $MX_2/CeO_2(111)$.

| Model | d (Å) | $E_{ad}$ (eV) | $E_{gap}$ (eV) | Bader charge (e) | |
|---|---|---|---|---|---|
| | | | | $MX_2$ | $CeO_2(111)$ |
| $MoS_2/CeO_2(111)$ | 2.84 | -1.31 | 1.51 | -0.029 | 0.029 |
| $MoSe_2/CeO_2(111)$ | 2.91 | -1.07 | 1.68 | -0.007 | 0.007 |
| $WS_2/CeO_2(111)$ | 2.76 | -1.36 | 1.67 | -0.028 | 0.028 |
| $WSe_2/CeO_2(111)$ | 2.90 | -1.12 | 1.72 | -0.002 | 0.002 |